# Polarized Neutron Measurement of Magnetic Order in $YBa_2Cu_3O_{6.45}$

H. A. Mook,[1] Pengcheng Dai,[1,2] S. M. Hayden,[3] A. Hiess,[4] S-H Lee,[5] and F. Dogan[6]

[1]Condensed Matter Sciences Division, Oak Ridge National Laboratory, Oak Ridge, TN 37831-6393

[2]Department of Physics and Astronomy, The University of Tennessee, Knoxville, TN 37996

[3]H. H. Wills Physics Laboratory, University of Bristol, Bristol BS8 1TL UK

[4]Institut Laue-Langevin, BP 156, 38042, Grenoble, France

[5]NIST Center for Neutron Research, National Institute of Standards and Technology, Gaithersburg, MD 20899

[6]Department of Ceramic Engineering, University of Missouri-Rolla, 222 McNutt Hall, 1870!Miner Circle, Rolla, MO 65409-0330

## Abstract

Orbital current order of the $d$-density-wave type (DDW) has been postulated to explain the pseudogap in the high temperature superconductors. We have performed neutron scattering experiments to search for this order and show here the results obtained on an $YBa_2Cu_3O_{6.45}$ sample using the best neutron spectrometers available. We argue that the data are consistent with a small, largely $c$-axis-directed moment, found below about 200K.

The microscopic origin of the pseudogap is one of the most puzzling attributes of the cuprate superconductors. A review of the various experimental techniques used to determine the pseudogap is given by Timusk and Statt.[1] It is generally considered to stem from pairing above the superconducting transition temperature $T_C$[2,3] or result from another ground-state differentiated from superconductivity by a quantum critical point.[4,5] Chakravarty $et\ al.$[6] have suggested that the pseudogap stems from a hidden order denoted by a $d$-density-wave (DDW) phase that competes with superconductivity and explains in a natural way the doping dependence of $T_C$. This state is similar to circulating currents around the Cu-O bonds resulting from the staggered flux phase of the $t$-$J$ model introduced by Hsu $et\ al.$[7] These currents would produce a small magnetic field directed largely along the $c$-axis of the superconducting crystal and should be visible in a sufficiently accurate neutron scattering experiment.[8] Here we present data demonstrating the present capability to observe such a state and show that the best measurements suggest the possibility of such a state rather than showing a null result.

An earlier neutron scattering experiment suggested that such a state occurs in $YBa_2Cu_3O_{6.6}$,[9] but a competing phase judged to stem from impurities made the identification of the DDW state difficult. The moments from the bond currents result in peaks at the ($h/2$, $k/2$, $l$) superlattice positions of the reciprocal lattice. This is the position that magnetism is found in the parent compound, $YBa_2Cu_3O_{6.15}$, which is an insulator with a Néel transition temperature well above room temperature and has been studied previously.[10,11] The defining attribute of the DDW state is that the moment should be largely along the $c$-axis. Small moments in the sample crystal originating from the parent compound should be located in the $a$–$b$ plane as is found for this material. Other magnetic impurity phases could be possible, but to our knowledge no such phases are known to have high temperature (above 50K) magnetic order along the $c$-axis for the YBa$_2$Cu$_3$O$_{6+x}$ system.

In order to determine the direction of the moments for the antiferromagnetic state it is necessary to use polarized neutrons, as outlined by Moon *et al*. in 1969.[12] The crystal used in the experiments is twinned so that **a*** could not be differentiated from **b***. We thus consider the crystal to be tetragonal so that the [1, 1, 0] direction is in the basal plane and the [1, -1, 0] direction perpendicular to the page is also in the basal plane. We describe our reciprocal lattice positions by **a***≈**b***=$2\Pi/(a+b)/2$ and **c***=$2\Pi/c$, and the scans are made in reciprocal lattice units (r.l.u.). Many of our results were obtained using the (1/2, 1/2, 1) reflection as shown in Fig. 1a. The momentum transfer **Q** for this reflection is determined by the difference in the incoming and outgoing neutron wavevectors, **k-k'**. Guide fields were used to provide a neutron polarization direction either vertical to the scattering plane (VF), or along **Q** (HF) and a standard spin-flipping coil was used in the scattered beam. All the data shown in the paper have been corrected for incomplete polarization. The magnetic neutron scattering intensity always originates from the moment projected on the plane perpendicular to **Q**. The polarization analysis technique can further determine how much of the moment lies along the direction of **Q$_\perp$**. We denote a scattering event in which the spin makes a 180° rotation as spin flip (SF), while scattering with no spin direction change is non-spin flip (NSF). For the HF case all magnetic scattering is SF and stems from all the moment in the crystal projected on the plane perpendicular to **Q**. The NSF scattering arises from nonmagnetic processes. For the VF case the magnetic scattering is divided up into SF and NSF parts. The SF part of the magnetic scattering lies in the scattering plane along **Q$_\perp$**. The NSF magnetic scattering stems from the moments pointing in the direction perpendicular to the scattering plane. A detailed discussion and the equations to determine moment direction from the HF and VF count rates are found in Refs. 9 and 12 so we will not reproduce them here. The moments from the DDW state are expected to be very small, and thus the signals observed very weak

The $YBa_2Cu_3O_{6.45}$ crystal used in the experiment weighed 25 g and had a $T_C$ of 55 K with a transition width of about 2.5 K. The method of growth incorporates the $Y_2BaCuO_5$ (the so-called green phase) into the crystal in small isolated pockets. Otherwise the sample is highly pure. Scans to observe Cu-O chain-order showed no particular type of order. This appears to be the best situation for the present experiment, as highly chain-ordered samples have shown magnetic impurity effects, possibly because of the ordered spaces available in the chains. The magnetic fluctuations for this crystal have been characterized previously.[13] Polarized beam data were first obtained at the NIST center for neutron research (NCNR) using the SPINS spectrometer. In this case supermirrors were used to both polarize and analyze the polarization of neutrons after scattering. Pyrolytic graphite was used as a monochromator and as a filter to remove higher order wavelength contamination in the beam. A flipping ratio of 8 was obtained and the data collected was corrected for the incomplete polarization of the beam using NSF scattering. A check of the (1, 1, 0) reflection showed no peak indicating that the higher order scattering from the monochromator was well removed. Figure 1b shows the result of a HF SF measurement through the (1/2, 1/2, 1) position at 20 K. The scan range on the high $Q$ side of the scan is limited by intense NSF scattering that is difficult to correct for with the rather small flipping ratio as seen in Fig 3a. Nevertheless, a peak is observed at (0.5, 0.5, 1) and a Gaussian fit results in a height of 11±2 counts with a width of 0.01±0.002 FWHM at the position 0.5±0.001 r.l.u. The peak is resolution limited, giving a correlation length of about 380Å. Figure 1c shows no evidence for a peak at a temperature of 150K. Multiple runs were averaged to obtain the observed error bars.

These scans took over two days each and are difficult to improve in a reasonable time period. Also, since the magnetic signal is so small, a higher flipping ratio is desired. The experiment was therefore moved to the IN20 spectrometer at the Institut Laue-Langevin neutron source in Grenoble, France.

The spectrometer arrangement of IN20 was the standard one used for polarization analysis and employed Heusler alloy crystals for the monochromator and analyzer. A flipping ratio of 19 was measured in both HF and VF field configurations. Open collimation was used from in front of the monochromator until after the analyzer. The neutron energy was 13.78 meV providing an energy resolution of about 1 meV FWHM, and three pyrolytic graphite filters were used to avoid higher order contamination. The (1/2, 1/2, 0) position was checked and no peak was found insuring that there was no second order contamination from the monochromator and analyzer.

For the case of present interest, the interpretation of the experiment is quite simple. Since we are looking for a predominately $c$-axis moment, we utilize the VF SF scan to determine if the moment is largely along $c$. The HF SF scan measures all the moment so that as long as the VF SF intensity is nearly as large as the HF SF intensity there is mostly $c$-axis order. This is the signature of the DDW state. Figure 1d shows a VF SF scan through the (0.5, 0.5, 1) reflection which should show the most intense signal from the orbital order. A large sloping background is found, but a peak is observed at (1/2, 1/2, 1) position. Least squares fitting of a Lorentization distribution finds a peak of 57±11 counts and at a position of 0.491±0.01 r.l.u. if the width is fixed at the resolution limited width of 0.0225 r.l.u. HWHM. It would be desirable to let the width vary in the fit, but the sloping background makes it impossible to get a reasonable value of the width in the fit. The NIST data suggests that the peak width is much smaller than the resolution on IN20. The resolution was determined from checks of the widths of the Bragg peaks of the sample. The second order of the (0.5, 0.5, 1) was also checked by removing the filters. The horizontal focusing makes the resolution quite broad with long tails. It, however, can be accounted for by a Lorentization distribution. The center of the distribution is at (0.5, 0.5, 1) within the error the measurement. The sloping background appears to depend only on the momentum transfer $Q$ so that it can be determined at positions in the reciprocal lattice away from (0.5, 0.5, l) where no magnetic signal is expected. It has been measured along the **c*** direction starting at positions outside of the range in (h, h, l) where any magnetic signal is observed. Figure 1e shows the result of subtracting the measured background fitted with a power law curve. The fit to the background is shown as a dashed line in Fig. 2b. The fit of the background subtracted data to a Lorentization distribution yields 54±5 counts for the height, 0.494±0.004 r.l.u. for the position, and 0.031±0.006 r.l.u. for the width, slightly larger than the resolution limited value.

Figure 2a shows the HF VF measurement. This must show a peak at least as large as the VF SF case as it contains all the magnetic scattering. The peak found at (0.5, 0.5, l) looks smaller, as the scale is somewhat different, but within the counting errors is the same. The resolution limited peak height is 37±10 counts with a peak position at 0.499±0.01 r.l.u. Since the VF SF case gives a signal as large as the HF SF case the moment must be largely directed in along the $c$-axis. Using the equations developed in Ref. 12 the results given in terms of counts would put the square of the c-axis moment strength at 73±13 and the square of the ab plane moment strength at -40±30. Given that the fitting procedure was identical in the two cases the argument for a largely c axis directed moment is a strong one while the ab plane moment is consistent with zero or a small number. It is difficult to provide an absolute moment value, but scaling from the Bragg intensities of our earlier measurements[9] would result in a moment value of about 0.0025$\mu_B$. The source of the large sloping background in the magnetic measurements is not clear. In Fig. 2b the SF HF and VF data have been subtracted. For a paramagnetic spin moment this should yield 1/2 the paramagnetic scattering. The error bars are large but the scattering does appear to decrease from about 50 counts to around 25 counts. This gives evidence that the background stems from some type of spin moment. The dashed line is the result from the measured background mentioned above. The background could stem from the paramagnetic scattering from the green phase although it appears to fall off with momentum rather quickly to stem from uncorrelated paramagnetic spins. The sloping background is not so visible in the NIST measurements, suggesting the open resolution captures more of the sloping background relative to the signal at the peak.

Figure 2c and 2d show the result measured at 70 K. Again, peaks are found on a sloping background for the VF SF and HF SF measurements. The fits are a Lorentzian distribution with the width fixed at the resolution limited values. The fits give heights of 34±12 and 57±14 counts for the VF and HF measurements and positions of 0.495±0.0005 and 0.501± 0.0004 r.l.u. The errors are too large to obtain a reliable moment direction, but are not inconsistent with the largely c-axis directed moment found at the lower temperature. The 70K data were not taken to distinguish a moment direction, but rather to determine if the magnetic signal persisted above T$c$. Figure 3a demonstrates the VF NSF scattering displays no peak at (1/2, 1/2 1) so there is no sign of a moment perpendicular to the scattering plane, although the count rate is sufficiently large that a small moment would be hard to observe. Most of the scattering stems from incoherent isotopic or spin sources or extra non-magnetic processes such as powder lines from impurity phases such as the $Y_2BaCuO_5$ or green phase. We have observed the magnetic scattering from the green phase if the sample is cooled to temperatures well below its ordering temperature of about 18 K. However, these disappear by 20 K, which is why this temperature was chosen for our lowest temperature runs with long counting times. The scattering on average in the NSF measurement is 5 times that for the SF case, which is the number observed in similar samples of different doping levels without extra impurities. The ratio for the $YBa_2Cu_3O_{6.6}$ sample that contained the magnetic impurity phase was nearly unity. Note that the small magnetic signals we are considering here would be completely unobservable in an unpolarized scan, as they would be added to the NSF scattering. Thus, while the unpolarized measurements by Stock *et al.*[14] saw no evidence for a magnetic peak, we are in agreement with their results.

Figure 3b shows scans made along the **c**\* direction at 70 K. A small peak is visible on the sloping background at (1/2, 1/2, 1) for both the HF SF and VF SF scans. This peak appears to be on the order of 50 counts as it should correspond to the scans in Fig. 2c and 2d. No sign of a peak is visible at (0.5, 0.5, 2), but this peak is expected to be smaller than the peak at (0.5, 0.5, 1) for the DDW state. It should be larger for *a–b* plane spin order as in the antiferromagnetic insulator. Figure 3c shows data for a scan HF SF scan at 170 K. There still appears to be some magnetic signal at (0.5, 0.5, 1) of about the same size as was found at 70 K. While a scan in the unpolarized mode would yield no information because of large non-magnetic scattering at different points in the scan, some information may be obtained about the magnetic peak intensity relative to the background. Figure 3d shows an unpolarized measurement of the intensity at (0.5, 0.5, 1) compared to that at (0.53, 0.53, 0). These positions are at the same total momentum transfer so that the two positions are obtained by turning the crystal. The measurement was made on the HB-1 spectrometer at Oak Ridge National Laboratory. This assures that the sloping background, which appears to be only a function of the total momentum, does not affect the measurements. The data are normalized by a long counting period at room temperature. There is obviously an increase in intensity at the (0.5, 0.5, 1) at low temperature. The difference seems to go to zero at around 200K. This would agree with the result in Fig. 1c, but appears different than Fig.!3c. It may be that the peak broadens at high temperatures, which would not affect the low-resolution results in Fig. 3c, but would result in a loss of peak height in the measurements of Fig.!1b and 3d. In any case, we are examining very small signals and it is difficult to establish definitive temperature dependence.

In summary, we have provided evidence for a small magnetic signal at the (0.5, 0.5, 1) position in $YBa_2Cu_3O_{6.45}$. The best interpretation of the data demonstrates that the magnetic moment is largely directed along the $c$-axis. This is different from the moment observed earlier for the parent compound or any known impurity phase with a peak at the observed lattice position. It is in the direction expected for the DDW phase. The exact temperature dependence is unknown, but the best evidence suggests the peak height of the signal is reduced in the neighborhood of 150K to 200K, which is on the scale of the pseudogap. Water in the sample would show up in the ratio of the background between the SF and NSF scattering since hydrogen has a huge spin incoherent scattering cross section. This ratio is around 1 to 5, which is the same for newly prepared samples. This makes it unlikely there is magnetism stemming from water.[15] Also water appears to enter the sample more easily when the chains are highly ordered leaving large free spaces in the chains. The moments cannot stem from a $c$-axis component of the sloping background found in the experiment as this does not peak at the magnetic lattice point. Better measurements would be desirable, but appear difficult at this time. Much longer counting times would be needed to materially affect the quality of the data. The best hope for improvement would seem to be the discovery of a sample that shows a larger effect. A number of samples have been investigated but highly underdoped samples show similar small effects, while highly-doped samples show no effect at all. As far as neutron scattering is concerned, the present results provide indications that orbital current phases are not ruled out and should continue to be seriously considered.

## Acknowledgements

We are grateful to S. Chakravarty for helpful discussions. This work was supported by the U.S. DOE under contract DE-AC0500OR22725 with UT-Battelle, LLC, and by U.S. NSF DMR-0139882.

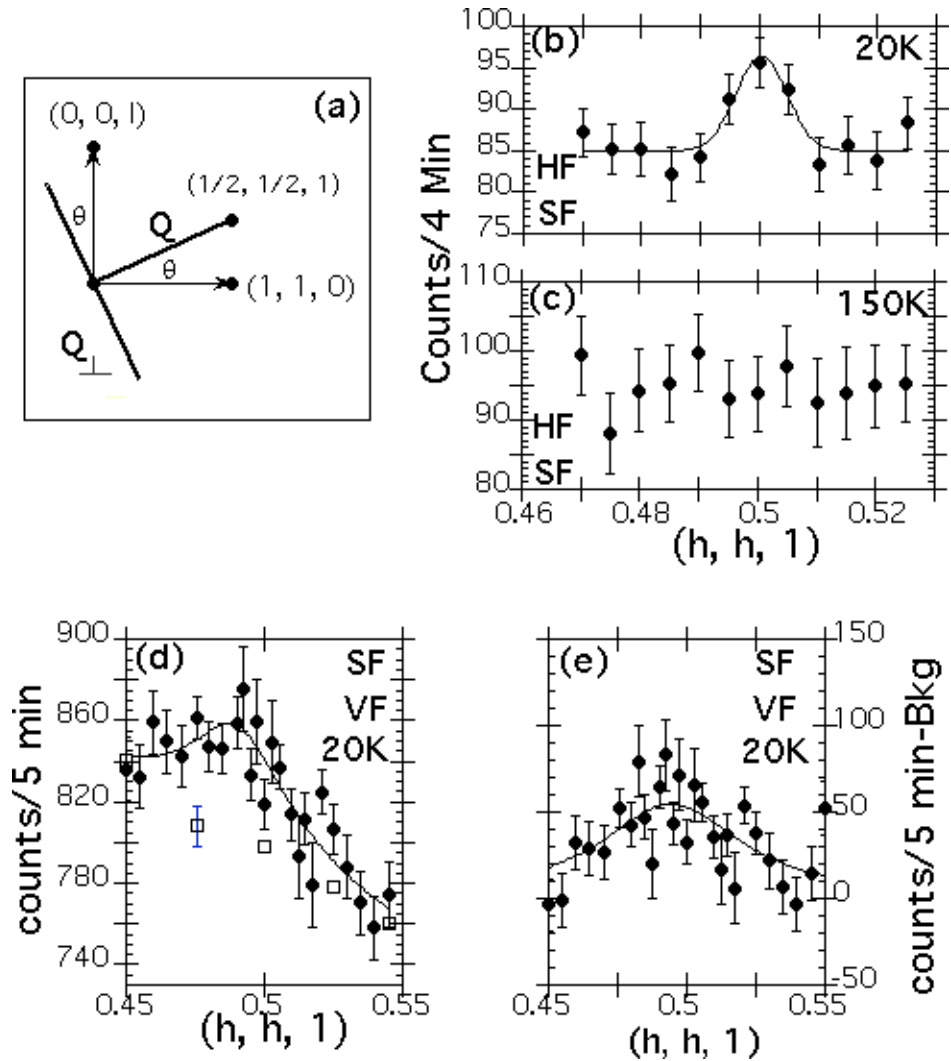

Fig. 1. Scattering diagram of the experiment and polarized beam results. (a) shows the scattering plane used in the experiment. (b) HF SF measurement of the magnetic order at 20 K around the (0.5, 0.5 1) reflection using the SPINS spectrometer. Several runs were averaged to obtain the errors shown. (c) SPINS spectrometer measurement at 150 K. (d) SF VF IN20 measurement at 20 K. A sloping background is found with the much coarser resolution used at IN20. The squares show a background measurement. (e) Same measurement as (d) but with sloping background as determined along **c**\* subtracted.

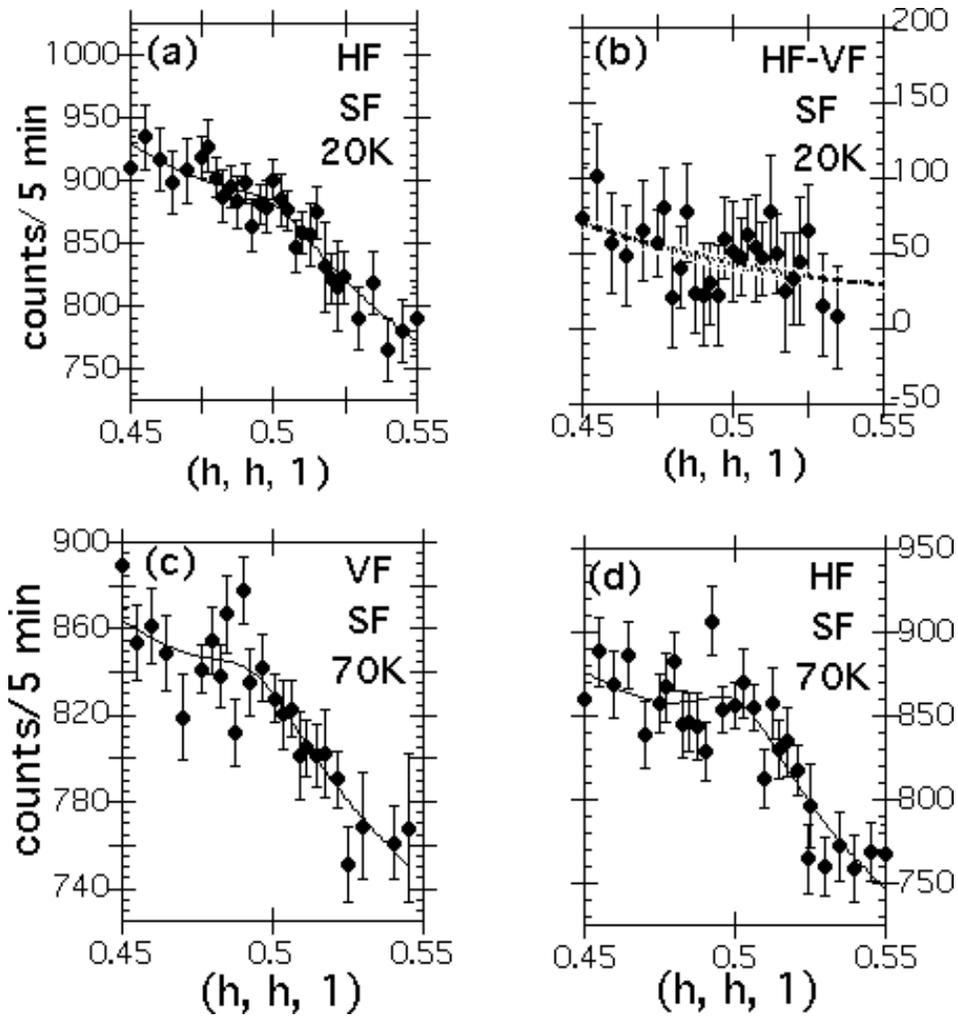

Fig. 2.  IN20 measurements of the magnetic scattering.  Multiple runs were averaged to obtain the observed error bars.  (a) shows the HF SF measurement at 20 K.  The background is even more steeply sloped than for the SF VF case.  (b) gives the difference of the SF, HF, and SF VF measurement.  For paramagnetic spin scattering this gives half the paramagnetic signal and should fall off as the magnetic form factor.  (c) SF VF measurement at 70 K.  (d) SF HF measurement at 70 K.

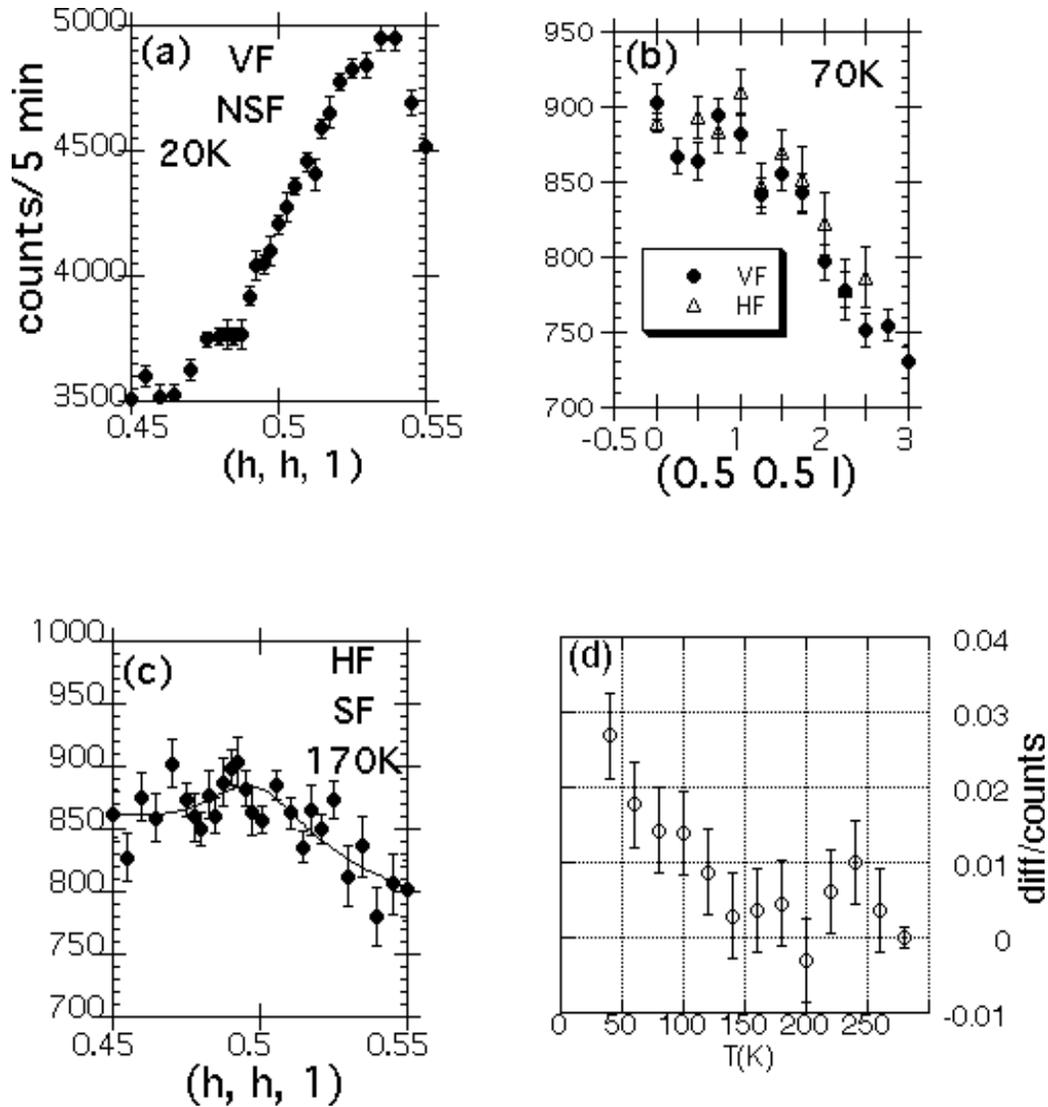

Fig. 3. IN20 polarized measurements and HFIR unpolarized determination of the temperature dependence of the magnetic scattering. (a) gives the VF NSF scattering. This largely stems from non-magnetic processes and is about 5 times larger than the total magnetic scattering. (b) SF, VF, and HF measurements for the **c**\* direction. (c) Shows the HF SF result for 170K. (d) gives an unpolarized measurement of the temperature dependence of the magnetic scattering. The result shown is obtained from the intensity at the (0.5, 0.5, 1) magnetic position, normalized to the background at (0.53, 0.53, 0).